\documentstyle[aps,epsfig,prb,multicol]{revtex}
\renewcommand{\vec}[1]{\mbox{\boldmath$#1$}}
\newcommand{\vecscr}[1]{\mbox{\boldmath$\scriptstyle #1$}}
\newcommand{\scr}{\scriptsize}
\newcommand{\etal}{\emph{et al.}}

\title{Quasiparticle Spectrum of $d$-wave Superconductors \\ in the Mixed
State: a Large Fermi-velocity Anisotropy Study}
\author{Luca\ Marinelli and B.\ I.\ Halperin \\ \small 
\emph{Physics Department, Harvard University, Cambridge, MA 02138}}
\date{August 21, 2001}

\begin{document}
\bibliographystyle{prsty}
\maketitle

\begin{abstract}
The quasiparticle spectrum of a two-dimensional $d-$wave
superconductor in the mixed state, $H_{c1} \ll H \ll H_{c2}$, is
studied for large values of the ``anisotropy ratio'' $\alpha_D =
v_F/v_\Delta$. For a square vortex lattice rotated by $45^\circ$ from
the quasiparticle anisotropy axes (and the usual choice of
Franz--Te\v{s}anovi\'c singular gauge transformation) we determine 
essential features
of the band structure asymptotically for large $\alpha_D$, using an
effective one-dimensional model, and compare
them to numerical calculations. We find that several features of the
band structure decay to zero exponentially fast for large
$\alpha_D$. Using a different choice of singular gauge transformation,
we obtain a different band structure, but still find qualitative
agreement between the 1D and full 2D calculations. Finally, we distort
the square lattice into a non-Bravais lattice. 
Both the one- and two-dimensional
numerical calculations of the energy spectra show a gap around
zero-energy, with our gauge choice, and the two excitation spectra 
agree reasonably well.      
\end{abstract} 


\begin{multicols}{2}

\section{Introduction}
The quasiparticle spectrum of $d-$wave superconductors
in the mixed state, $H_{c1} \ll H \ll H_{c2}$, has been the object of
recent detailed studies. High-$T_c$ superconductors are extreme
Type-II superconductors and in a magnetic field $H \gg H_{c1}$ develop
a vortex lattice, the geometry of which depends somewhat on the
strength of the magnetic field. The $d-$wave symmetry implies the existence
of four points on the zero-field Fermi surface where the gap
vanishes. This fact has important implications for the low-temperature
thermodynamics of these systems, as there are states available even at
very low temperatures.

The main questions that have been addressed in the literature are the
fate of these gapless points in a finite magnetic field and, more
generally, the nature of low-energy states of the quasiparticle
spectrum. The standard approach to studying the quasiparticle spectrum
in a superconductor is through the Bogoliubov--de Gennes equation
\cite{degennes89}, which allows for a spatially varying order
parameter in a natural way.  
Gor'kov and Schrieffer \cite{gorkov98} and, in a more recent
paper, Anderson \cite{anderson98} neglected in a first
approximation the spatially dependent superfluid velocity in the
vortex lattice. They predicted that the quasiparticle spectrum in a
magnetic field $H \ll H_{c2}$ is characterized by broadened Landau
levels. However, their key assumption, namely treating the superfluid 
velocity in perturbation theory, is questionable
as work by Mel'nikov \cite{melnikov99} shows that this singular
perturbation strongly mixes Landau levels.

A new approach to tackle these questions was pioneered by Franz and
Te\v{s}anovi\'{c} \cite{franz00}. They introduced a singular gauge
transformation that takes into account the supercurrent distribution
and the magnetic field on an equal footing. Starting from the
linearized Bogoliubov--de Gennes equation \cite{simon97} in a magnetic
field, they mapped 
the original problem onto one of diagonalizing a Dirac Hamiltonian in
an effective periodic vector and scalar potential with vanishing
magnetic flux in the unit cell. The new Hamiltonian is more suited to
numerical diagonalization, using standard band structure
calculation techniques. Within their model, Franz and Te\v{s}anovi\'{c} found that the
low-energy quasiparticle spectrum stays gapless at the center of the
Brillouin zone ($\Gamma$ point), even in the presence
of a magnetic field. They also noticed that, for large
anisotropy ratio $\alpha_D = v_F/v_\Delta$ (where $v_\Delta =
\Delta_0/p_F$ is the quasiparticle velocity parallel to the Fermi
surface and $\Delta_0$ is the maximum magnitude of the
bulk zero-field gap function), more low-energy states arise. For
larger values of $\alpha_D$, they found entire lines in the Brillouin
zone where the energies appeared to vanish, within the uncertainty of
their numerical calculations.

Because it is difficult to decide by direct numerical calculations whether
there are actual zeroes of the dispersion relations, we undertook in a
previous paper \cite{marinelli00} to employ symmetry considerations,
as well as a
perturbative analysis and alternative calculational methods to
elucidate the properties of solutions of the Franz--Te\v{s}anovi\'{c}
(FT) equations. The present work continues these efforts by analysing a
one-dimensional variant of the FT model, which was
also considered by Knapp, Kallin and Berlinsky \cite{knapp00}.

The one-dimensional model has the advantage that it can be solved
numerically or analytically with arbitrary accuracy, and its
asymptotic behavior, for large values of $\alpha_D$ can be readily
extracted. Moreover, the behavior of the 1D model and the
two-dimensional FT model are very close to each
other for large values of the anisotropy ratio $\alpha_D$. Thus we can
gain valuable insight into the solutions of these equations.

We note that relatively large values of the anisotropy ratio are relevant to
actual high-$T_C$ superconductors, namely $\alpha_D$ ranging between
10 and 20. (From angle-resolved photoemission spectroscopy
and thermal conductivity measurements, the value of $\alpha_D$ for
high-$T_c$ superconductors turns out to be about 14 for YBCO
\cite{chiao99} and 20 for Bi2212 \cite{chiao99,mesot99}.)

The singular gauge transformation introduced by Franz and
Te\v{s}anovi\'{c} is exact in the region of space outside the vortex core
where the magnitude of the gap parameter $|\Delta(\vec{r})|$ is
independent of position. As we are interested in ther situation of
weak magnetic fields, where the vortex cores occupy a very small
fraction of the total area, it is natural to suppose that errors in
the energies introduced by an arbitrary treatment of the core interior
would not be significant. However, the situation is more
subtle, as was noted by Vafek, Melikyan and Te\v{s}anovi\'{c}
\cite{vafek01}.

\begin{figure}
\noindent
\epsfxsize=8.5cm
\epsffile{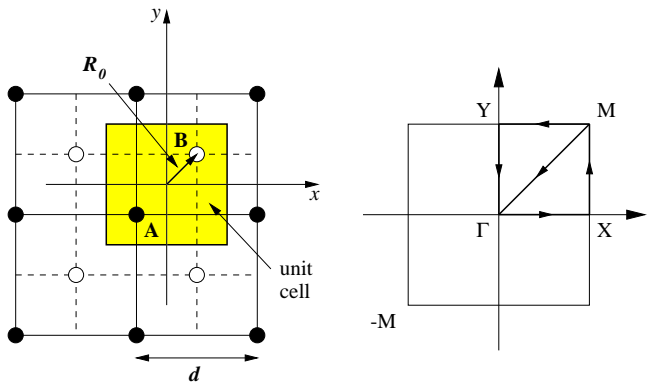}
\caption{(a) A and B sublattices and vortex lattice unit cell. (b) The
corresponding magnetic Brillouin zone.}
\label{figAB}
\end{figure} 

The energy states of interest have a large weight in the vicinity of
the vortex cores, and the energies can be shifted by amounts which are
finite on the scale of $E_1 = \hbar v_F/d$ ($d$ is the distance
between nearest vortices belonging to the same FT sublattice, as defined
in Fig.~\ref{figAB}), depending on what one assumes for the behavior
inside the vortex core. Alternatively, one can introduce a
mathematical boundary condition at the radius of the vortex core. The
wavefunction, which obeys the FT equations
outside the core, must satisfy appropriate boundary conditions at the
boundary of the core. It appears, however, that there are a variety of
possible boundary conditions which remain mathematically sensible and
distinct in the limit of vanishing core radius. 
An analysis of the behavior inside the core is therefore ultimately
necessary to find the behavior appropriate to a given microscopic
Hamiltonian or for an actual high-$T_C$ superconductor, even in the
limit of weak magnetic fields. 

We do not address this problem here. Instead, we follow the approach
of Franz and Te\v{s}anovi\'{c} \cite{franz00} and our own subsequent work 
\cite{marinelli00}, in which, for a given choice of the singular gauge
transformation, the vortex core is treated by smoothing the superfluid
velocities over a small area. Although this approach appears to be
well defined in the limit where the core-radius vanishes relative to
the vortex separation, the energy spectrum obtained does depend on the
particular choice made in the FT gauge
transformation. (\emph{i.e.}, the energy spectrum depends on which
vortices are assigned to each of the two FT sublattices.) 
Similarly, each of the one-dimensional models considered in the
present paper depend on the choice of FT gauge used
in the 2D model from which it was derived, and the resulting
dispersion relations reflect the differences between the parent 2D
models. 

The essence of the one-dimensional model is to smear the vortices in the $y$ 
direction (as defined in Fig.~\ref{figAB}). This direction is the
``hard'' direction, meaning that wavefunctions in this direction
change very slowly. In particular, if we focus on the
low-energy states, any fluctuation of the quasiparticle wavefunctions
in the $y$ direction (for large anisotropy) will be energetically
very costly. For this reason, it is reasonable to assume that
asymptotically for $\alpha_D \to \infty$ the lowest-energy states will
be translationally invariant in the $y$ direction (and, more generally,
they will be represented by plane waves in the $y$ direction with a
wavevector $k_y$ equal to a reciprocal vector of the vortex lattice, if
higher energy states are to be taken into account). 
Of course, this assumption depends crucially on the
orientation of the vortex lattice with respect to the quasiparticle
anisotropy axes. We consider a square vortex lattice tilted by
$45^\circ$ with respect to the anisotropy axes. If the angle was
different, a line going through a vortex along the ``hard'' direction
would not necessarily intersect other vortices (incommensurate case) or
could intersect another vortex some number of unit cells away from the
vortex we considered (commensurate case). In either case the
description would be more complicated than the one considered here.

Mel'nikov \cite{melnikov99} was the first to realize that the
Bogoliubov--de Gennes equation for a $d-$wave superconductor in the
mixed state is simplified in the large anisotropy limit $\alpha_D \gg
1$. He derived a one-dimensional approximate model and described how
to solve it, although he confined his analysis to the semiclassical
version of these solutions. Knapp, Kallin and Berlinksy \cite{knapp00},
starting from the FT equations, derived a 1D model valid for large
$\alpha_D$ and carried out a fully quantum mechanical calculation of the properties
of this model. They worked in the singular gauge introduced by Franz
and Te\v{s}anovi\'{c} \cite{franz00} matched the solutions in terms
of parabolic cylinder functions at the boundaries of the unit cell to
obtain an exact excitation spectrum for the one-dimensional
Hamiltonian. This matching, although exact, was nevertheless carried 
out numerically and therefore did not give analytical 
information on the dependence of the features of  the band structure 
on the anisotropy $\alpha_D$. 

We have found that the most important features of the
one-dimensional model can be understood and calculated using an
approach which emphasizes the underlying physics in a more transparent
way. In essence, the low-energy states are localized around one of the
two vortices in the unit cell and, in the large anisotropy limit, 
the overlap between linearly independent wavefunctions can become 
exponentially small as a function of $\alpha_D$ if
the potential wells these states are confined in are spatially separated. In
the singular FT gauges that exhibit this exponential behavior,
features of the energy bands, such as the minimum energy in the
$\Gamma$Y direction (away from the $\Gamma$ point) or the bandwidth of
the $\Gamma$X lowest energy band, can be calculated analytically in the WKB approximation. 

\begin{figure}
\noindent
\epsfxsize=8.5cm
\epsffile{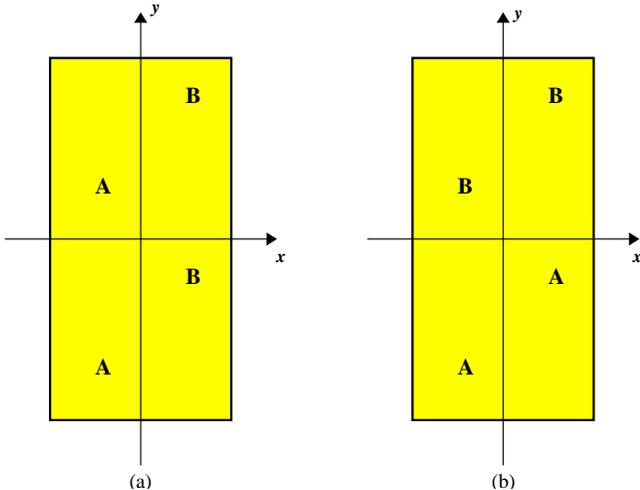}
\caption{(a) $ABAB$ vortex lattice unit cell (plotted with four vortices
per unit cell). (b) $AABB$ vortex lattice unit cell.}
\label{fig:lattice}
\end{figure} 
 
The lower dimensionality of the model under study in this paper,
allows also for much smaller scale numerical calculations to find the
quasiparticle spectrum than the two-dimensional case,
as was already pointed out by Knapp, Kallin and Berlinsky
\cite{knapp00}. Some of the features which were hard to resolve in the full 
two-dimensional problem, such as the band structure along the $\Gamma$X
symmetry line and the apparent lines of zero-energy states that formed
parallel to the $\Gamma$X line and intersected the $\Gamma$Y line at
non-symmetry points, become readily accessible in the one-dimensional
model. In this work we have performed such numerical calculations for
two different singular gauge choices and found good agreement with the
asymptotic analysis and the data available for the two-dimensional
model in both cases. In the first singular gauge, which we called $ABAB$
following Vafek \etal \cite{vafek00} (see Fig.~\ref{fig:lattice}), the
low-energy features of the spectrum exhibit the above-mentioned exponential
dependence on $\alpha_D$, while in the second gauge, which we called $AABB$,
they do not. 

If, after including the appropriate vortex core physics, the correct
energy spectrum in the continuum limit turns out to be related to the
results in the $ABAB$ gauge, we could use the above results to improve
our understanding of the relation between the semiclassical
description of the quasiparticle spectrum in the vortex state, as
derived by Volovik \cite{volovik93} and the fully quantum-mechanical
picture. The semiclassical density of
states is characterized by two crossover energy scales, $E_1$ and $E_2$ ($E_2
<E_1$). For energies $E \gg E_1 = \hbar v_F/d$ the density of states
is, on average, quantitatively identical to that in the absence of a magnetic
field, while for $E<E_1$ the density of states can be 
evaluated in a semiclassical approximation and turns out to be
constant. The semiclassical description breaks down at the energy
scale $E_2$, where a full quantum mechanical calculation becomes
necessary. In our previous paper \cite{marinelli00}, we noticed that
the value of the crossover energy scale $E_2$ predicted by Kopnin and
Volovik \cite{kopnin96,volovik97,volovik97b} $E_2^{KV} \approx \hbar
v_\Delta / d$ did not seem to match with our numerical findings, at
least for the vortex lattice geometries we considered. In
fact, it seemed that $E_2$ went to zero much faster than linearly in
$1/\alpha_D$. Furthermore, recent specific heat \cite{revaz98} and
low temperature thermal conductivity \cite{chiao99_2}
experiments have been performed in the regime $E < E_2^{KV}$ and still
show good agreement with the semiclassical predictions, which 
seems to suggest that the correct crossover scale between the quantum
mechanical and semiclassical regimes is much smaller than $E_2^{KV}$,
for large anisotropy ratios. The one-dimensional model
provides the key to this puzzle, as both the bandwidth of
the lowest $\Gamma$X energy band and the minima of the $\Gamma$Y
band (away from the $\Gamma$ point) are shown to be suppressed exponentially
for large $\alpha_D$, instead of linearly. In particular,  
the larger of these two energies will determine the crossover between
the semiclassical and quantum mechanical regimes, where the details of
the band structure become important to determine the density of states. 

We also studied the one-dimensional energy spectrum of a non-Bravais
vortex lattice, with the usual choice of singular gauge ($ABAB$). 
In this case, we previously found \cite{marinelli00} that the
two-dimensional band structure becomes gapped. It has been suggested
recently by Vishwanath \cite{ashvin01} that this gap may be a spurious
feature of the Franz--Te\v{s}anovi\'{c} $ABAB$ model; indeed one can
make other choices for the singular gauge transformation, assigning
the vortices to sublattices in a different way, so that the spectrum
is gapless for the very geometry in question. We do not try to answer
here the question of whether a gap would be present in this geometry
when one treats properly the region inside the vortex cores. We limit
ourselves here to the observation that given a particular choice of
the singular gauge transformation the spectra of the two-dimensional
model and of the one-dimensional model derived from it agree
reasonably well at long values of $\alpha_D$. In both cases the lowest positive
and negative energy bands are separated by a finite energy gap, and
the minimum energy separation between the two bands does not  occur at
the $\Gamma$ point (for anisotropy $\alpha_D \neq 1$). Similarly to
other results discussed in this paper, this energy gap could be a
feature which depends on the choice of vortex core physics. Whether it
survives or not the correct choice of boundary conditions is not
within the scope of this paper. We limit ourselves to note that given
a singular gauge transformation, we find good agreement between the
spectra of the one- and two-dimensional Hamiltonians. 

To summarize the structure of the paper, the one-dimensional model
Hamiltonian is derived in Sec.~\ref{sect:TI}. The asymptotic analysis of the
band structure for large anisotropy follows in Sec.~\ref{sect:asymptotics}. 
Numerical results can be found in Sec.~\ref{sect:numerics}, where
a comparison between two different singular gauge choices is shown
explicitly, together with a study of the excitation spectrum of a
non-Bravais vortex lattice. Conclusions follow in
Sec.~\ref{sect:conclusion}. A derivation of the WKB approximation
and an account of the numerical methods can be found in Appedix A and
B, respectively.

\section{Derivation of the One-Dimensional Model}
\label{sect:TI}
The properties of the quasiparticle spectrum in a superconductor in an
external magnetic field are best studied in the framework of the
Bogoliubov--de Gennes equation \cite{degennes89}. To establish
notations and for overall clarity, we will very briefly summarize the most 
important points of the derivation of the Bogoliubov--de Gennes equation 
for a $d$-wave superconductor in the mixed  state, more details con be 
found in our earlier paper \cite{marinelli00} (see also \cite{vafek00}).   

We will consider the Bogoliubov--de Gennes operator for a 
$d_{xy}$-superconductor 
instead of the more conventional $d_{x^2-y^2}$, purely for notational
simplicity; results do not depend on this choice. Also,
because we are only interested in the low-energy properties of the
spectrum and because we consider magnetic fields $H \ll H_{c2}$
(i.e. such that the size of the vortex lattice unit cell is much
larger than $1/k_F$), we 
can linearize the Bogoliubov--de Gennes equation around one of the
four gapless points on the Fermi surface. Choosing $\vec{p} =(0,p_F)$,
the linearized Bogoliubov--de Gennes operator reads \cite{simon97} 
\begin{equation}
\label{bdglin}
\tilde{\cal H}_{\mbox{\scr lin}} = \left(
\begin{array}{cc}
v_F (p_y -\frac{e}{c} A_y) & \frac{1}{p_F} \{p_x, \Delta(\vec{r})\} \\
\frac{1}{p_F} \{p_x, \Delta^*(\vec{r})\} & -v_F (p_y+\frac{e}{c} A_y)
\end{array}
\right),
\end{equation}
where $\Delta(\vec{r}) = \Delta_0 e^{i \phi(\vecscr{r})}$ is the
Ginzburg--Landau order parameter and the brackets represent
symmetrization $\{a,b\} = \frac{1}{2} (ab+ba)$.

In the singular gauge introduced by Franz and Te\v{s}anovi\'{c} 
\cite{franz00}, we can eliminate the position dependent phase factor 
$e^{i \phi(\vecscr{r})}$ from the off-diagonal components of the
Bogoliubov--de Gennes equation (\ref{bdglin}). The resulting
Bogoliubov--de Gennes operator after the gauge transformation is
\begin{eqnarray}
{\cal H}_{\mbox{\scr lin}} &=& \left(
\begin{array}{cc}
v_F p_y & v_\Delta p_x \\
v_\Delta p_x & - v_F p_y
\end{array}
\right) \nonumber \\
&+& m \left(
\begin{array}{cc}
v_F v^A_{sy} & \frac{v_\Delta}{2} (v^A_{sx}-v^B_{sx}) \\
\frac{v_\Delta}{2} (v^A_{sx}-v^B_{sx}) & v_F v^B_{sy}  
\end{array}
\right),
\label{Htransf}
\end{eqnarray}
where $v_\Delta = \Delta_0/p_F$. The superfluid velocities
corresponding to the $A$ and $B$ sublattices of the vortex lattice, as
defined in Fig.~\ref{figAB}, are
\begin{equation}
\vec{v}^\mu_s = \frac{1}{m} (\hbar \nabla \phi_\mu - \frac{e}{c}
\vec{A}), \ \  \mu = A, B
\end{equation}
where $\phi(\vec{r}) = \phi_A(\vec{r})+\phi_B(\vec{r})$. 
The operator (\ref{Htransf}) describes the dynamics of a non-interacting Dirac
particle in a periodic scalar and vector potential and general features of its 
spectrum have been studied in detail elsewhere
\cite{franz00,marinelli00,vafek00}.   

The Bogoliubov--de Gennes equation should in principle be solved
self-consistently, thus finding the order parameter distribution (from
which the geometry of the minimum free-energy vortex
lattice configuration can be inferred) together with the energies and 
wavefunctions of the quasiparticle excitations. At low temperatures
and for the magnetic field range considered here, to a good
approximation, the order parameter
distribution can be computed in Ginzburg--Landau theory and
substituted in the Bogoliubov-de Gennes equation, without iterating it
to achieve self-consistency. We studied the quasiparticle spectrum in
this approximation, choosing a square vortex lattice rotated by
$45^\circ$ from the quasiparticle anisotropy axes. 

In this work we want to focus on the large anisotropy  $\alpha_D = 
v_F/v_\Delta \gg 1$ limit. For this reason we have decided to study 
a one-dimensional model which
captures the essential physics of the large-anisotropy regime and we
believe becomes asymptotically exact in the $\alpha_D \to \infty$ and
low-energy limit.
The essence of the model is to smear out the vortices in the $y$
direction (the ``hard'' direction). The absolute value of the
low-energy two-dimensional
wavefunctions has only small ripples and overall fluctuations in the
``hard'' direction, thus motivating the claim that asymptotically
the low-energy eigenstates of (\ref{Htransf}) should factor into a plane
wave in the $y$ direction times a Bloch state in the $x$ direction.
 
To simplify equations we will write lengths in units of
the distance between nearest-neighbor vortices $d$ and energies in
units of $E_1=\hbar v_F/d$. Let us define the dimensionless periodic potentials
\begin{equation}
U^{A,B}_{x,y} = \frac{m d}{\hbar} v^{A,B}_{sx,y} = \partial_{x,y}
\phi^{A,B} - \frac{e d}{\hbar c} A_{x,y}
\end{equation}
where the vector potential is chosen to satisfy the Landau gauge $A_x =
-d B y$ and $A_y =0$. The Bogoliubov--de Gennes operator can then be
rewritten in dimensionless form
\end{multicols}
\begin{equation}
{\cal H} = \left(
\begin{array}{cc}
\frac{1}{i} \partial_y + U^A_y & \frac{1}{i \alpha_D} \partial_x +
\frac{1}{2 \alpha_D} \left(U^A_x-U^B_x\right) \\
\frac{1}{i \alpha_D} \partial_x + \frac{1}{2 \alpha_D}
\left(U^A_x-U^B_x\right)& - \frac{1}{i} \partial_y +U^B_y
\end{array}
\right).
\label{Hdimless}
\end{equation} 

\begin{multicols}{2}
The one-dimensional approximation amounts to assuming that gauge
invariant quantities can only be a function of $x$. In particular,  
the potentials $U^{A,B}_y$ and $U^{A,B}_x$ are gauge
invariant quantities and thus
\begin{equation}
\partial_y \left(\partial_x \phi^{A,B} + 2 \pi y \right) =0
\end{equation}
which, away from the vortices, implies 
\begin{equation}
\partial_y \phi^{A,B} = - 2 \pi x + \mbox{const.}
\end{equation}
The constant of integration in the above expression changes
discontinuously when crossing a line of vortices and can be determined
by imposing that the potentials
$U^{A,B}_y$ are periodic (with period 1) and have zero average and that
\begin{equation}
\oint_{\cal C} \nabla \phi^{A,B} \cdot d\vec{l} = 2 \pi N_{A,B},
\end{equation}
where $N_{A,B}$ is the number of $A$ or $B$ vortices in the unit cell.
The path $\cal C$ is a rectangle enclosing a line of vortices with
infinitesimal width in the $x$ direction and spanning the vortex
lattice unit cell in the $y$ direction. 
Furthermore, in the large-anisotropy limit, $v_{sx}^A$ and $v_{sx}^B$ 
will both vanish.
The periodicity of the potentials can be used to further simplify the
calculation by writing the wavefunctions in Bloch form 
\begin{equation}
\left(
\begin{array}{c}
v_{\vecscr{k}}(x,y) \\ w_{\vecscr{k}}(x,y)
\end{array} \right) = e^{i (k_x x + k_y y)} \left(
\begin{array}{c}
V_{\vecscr{k}}(x) \\ W_{\vecscr{k}}(x)
\end{array} \right).
\end{equation}
The final result is that the one-dimensional effective Hamiltonian
acting on Bloch states takes the form 
\begin{equation}
{\cal H}_{\mbox{\scr 1D}} = \left(
\begin{array}{cc}
k_y + U^A(x) & \frac{1}{i \alpha_D} \partial_x +\frac{1}{\alpha_D} k_x\\
\frac{1}{i \alpha_D} \partial_x +\frac{1}{\alpha_D} k_x & -k_y +U^B(x)
\end{array}
\right).
\label{H1D}
\end{equation}

We have considered three different vortex arrangements, two of which
represent the same physical situation of a square Bravais
lattice of vortices and one is an inversion-symmetric non-Bravais lattice.
The two Bravais vortex lattices differ in the labeling of the
vortices, as can be seen in Fig.~\ref{fig:lattice}. Following Vafek
\etal \cite{vafek00}, we will refer to them as the $ABAB$ and $AABB$
configuration, respectively. 

\begin{figure}
\noindent
\epsfxsize=8.5cm
\epsffile{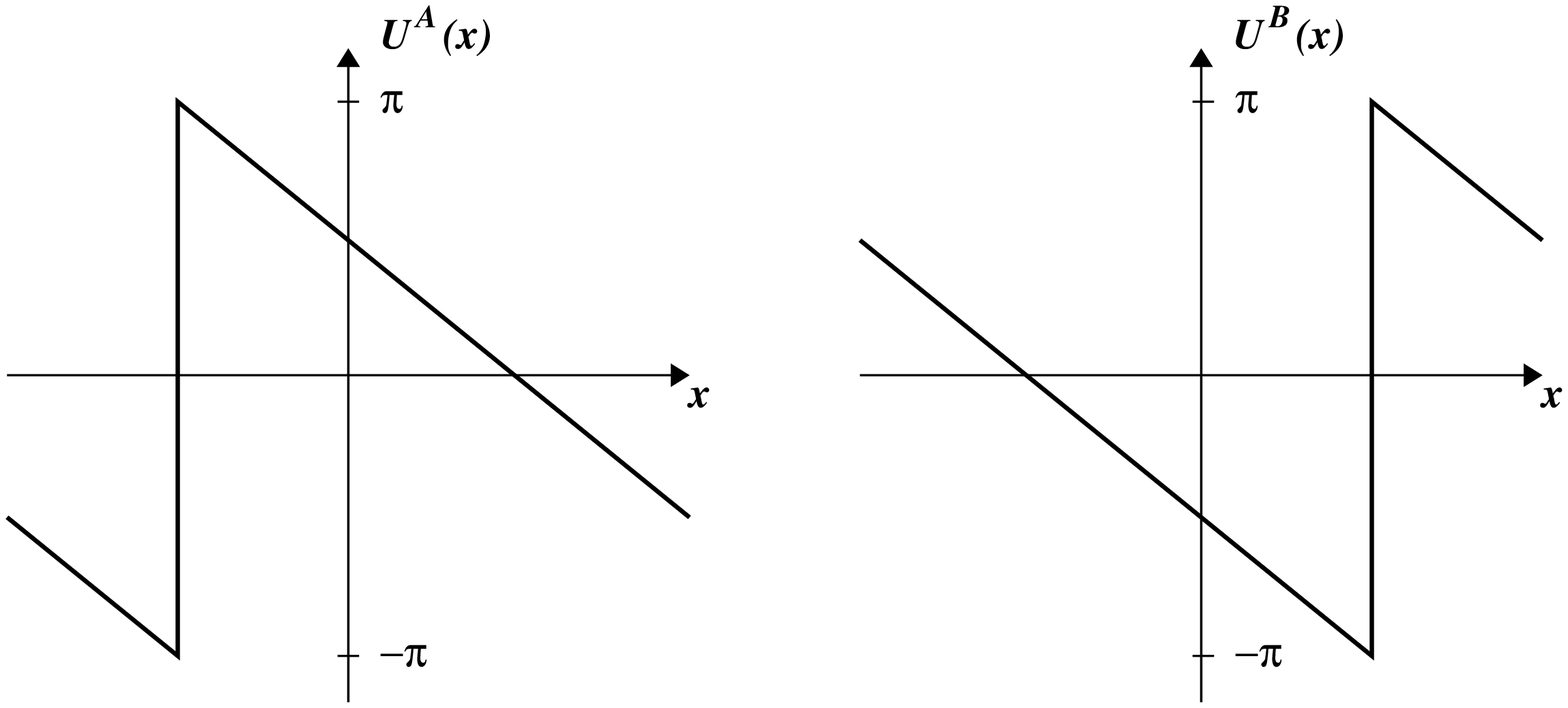}
\caption{The one-dimensional periodic potentials $U^A(x)$ and $U^B(x)$ in the
$ABAB$ gauge are plotted in the unit cell. Energies are in units of
$\hbar v_F/d$.}  
\label{figU}
\end{figure} 

In the $ABAB$ case, we can work with a unit cell with one $A$ and one
$B$ vortex. In this case the potentials $U^A(x)$ and $U^B(x)$ can be
calculated as outlined above to find
\begin{equation}
U^A(x) = \left\{
\begin{array}{cc}
-2 \pi x - \frac{3}{2} \pi, & -\frac{1}{2} \leq x \leq -\frac{1}{4} \\
-2 \pi x + \frac{1}{2} \pi, & -\frac{1}{4} < x \leq \frac{1}{2} 
\end{array} \right.
\end{equation}
and 
\begin{equation}
U^B(x) = \left\{
\begin{array}{cc}
-2 \pi x - \frac{1}{2} \pi, & -\frac{1}{2} \leq x \leq \frac{1}{4} \\
-2 \pi x + \frac{3}{2} \pi, & \frac{1}{4} < x \leq \frac{1}{2}. 
\end{array} \right.
\label{eq:ABAB}
\end{equation}
The $U^A(x)$ and $U^B(x)$ potentials in the $ABAB$ gauge are plotted
in Fig.~\ref{figU}.

Alternatively, we considered the $AABB$ configuration, in which case one has to
work with a unit cell containing four vortices (two of type $A$ and
two of type $B$). In this case the discontinuity of the potentials
across a line of vortices is $\pi$ (this should be compared to $2 \pi$ in the
case of the $ABAB$ lattice) and the $U^A(x)$ and $U^B(x)$ potentials are
identical across the unit cell
\begin{equation}
U^A(x) = U^B(x) = \left\{
\begin{array}{cc}
-2 \pi x - \pi, & -\frac{1}{2} \leq x \leq -\frac{1}{4} \\
-2 \pi x, & -\frac{1}{4} \leq x \leq \frac{1}{4} \\
-2 \pi x + \pi, & \frac{1}{4} \leq x \leq \frac{1}{2}. 
\end{array} \right.
\label{eq:AABB}
\end{equation}

The one-dimensional model can be derived also for a class of
non-Bravais lattices. In particular, we studied an inversion-symmetric
lattice, whose quasiparticle spectrum we analyzed in detail in a
previous paper \cite{marinelli00}. This lattice has two vortices per
unit cell whose distance along the diagonal, $2 R_0$ (see
Fig.~\ref{figAB}), is different from
$(\sqrt{2}/2) d$ (which corresponds to evenly spaced $A$ and $B$
vortices, necessary condition for a Bravais lattice). For
concreteness, let us consider the case where the two
sublattices $A$ and $B$ are still square lattices with spacing $d$,
but the distance between nearest $A$ and $B$ vortices is now $(2 x_v
\sqrt{2}) d$. Switching back to dimensionless units, the $A$ and $B$
vortices in the unit cell are located 
at coordinates $(-x_v,-x_v)$ and $(x_v,x_v)$ respectively, instead of
$(\pm 1/4,\pm 1/4)$, as in the Bravais lattice case.
Here we are particularly interested in determining whether 
the appearance of a gap in the excitation spectrum (as was found
before in the two-dimensional numerical analysis) is peculiar to the
two-dimensional model or is a feature of the one-dimensional
approximation as well. To this end, we have calculated the $U^A(x)$
and $U^B(x)$ potentials when the lines of vortices are located at
$x=\pm x_v$ 
\begin{equation}
U^A(x) = \left\{
\begin{array}{cc}
-2 \pi x - (2 x_v +1) \pi, & -\frac{1}{2} \leq x \leq -x_v \\
-2 \pi x - (2 x_v -1) \pi, & -x_v < x \leq \frac{1}{2} 
\end{array} \right.
\label{eq:xvA}
\end{equation}
and 
\begin{equation}
U^B(x) = \left\{
\begin{array}{cc}
-2 \pi x + (2 x_v -1) \pi, & -\frac{1}{2} \leq x \leq x_v \\
-2 \pi x + (2 x_v +1) \pi, & x_v < x \leq \frac{1}{2}. 
\end{array} \right.
\label{eq:xvB}
\end{equation}
Note that the discontinuity of the $U^A(x)$ and $U^B(x)$ potentials at
the vortex lines is always $2 \pi$, regardless of the location of the vortices.
These potentials will be used in the analysis of the excitation spectrum,
both in numerical calculations and in the asymptotic analysis of
prominent features of the band structure for the $ABAB$ vortex lattice.

\section{Asymptotic analysis of the band structure for the $ABAB$ vortex
lattice}
\label{sect:asymptotics}
The one-dimensional model can be solved exactly by
matching the parabolic cylinder function solutions \cite{melnikov99}
across the boundary of the unit cell. This method requires a numerical
exact calculation to match the boundary conditions and leads to
results which are practically indistinguishable from the numerical
diagonalization of the one-dimensional model, as shown by Knapp,
Kallin and Berlinsky \cite{knapp00}. 
Here we use an alternative approach which does not provide an analytical
solution of the one-dimensional model for arbitrary values of the
anisotropy but it allows us to understand in a relatively simple way
the leading behavior of the
quasiparticle spectrum for large values of the anisotropy ratio
$\alpha_D$. 
In this section we will limit our analysis to the $ABAB$ vortex lattice,
although we will comment on the other gauge choice that was introduced
in the previous section. 

In essence, in the large anisotropy limit and for the $ABAB$
configuration, the low-energy wavefunctions are
localized in the triangular potential wells $U^A(x) +k_y$ and
$U^B(x)-k_y$, defined in equation (\ref{H1D}) with the potentials
(\ref{eq:ABAB}). The tails of these wavefunctions overlap in
the classically forbidden regions of the potential wells, and the
overlap becomes exponentially small for large $\alpha_D$. 
Using $1/\alpha_D$ as our ``small'' parameter, we
can develop a WKB approach which can be used to calculate the
leading exponential dependence on $\alpha_D$ of the tails of the
wavefunctions. This, in turn, allows us to understand the features of
the band structure, like the lowest energy band widths and energy
splittings at near crossings. 

\begin{figure}
\noindent
\epsfxsize=8.5cm
\epsffile{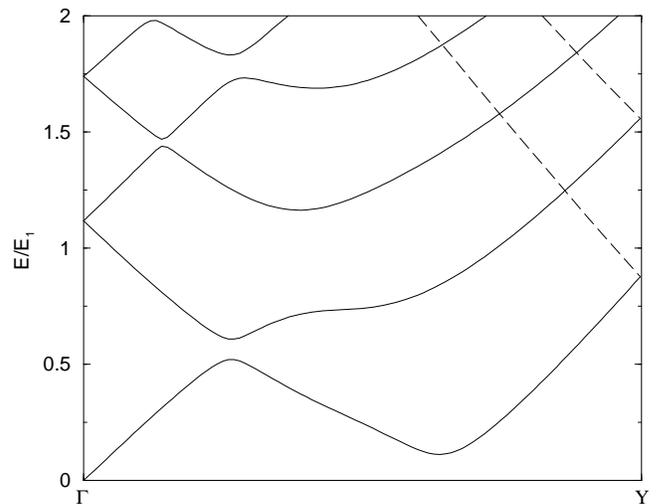}
\caption{Band structure of the one-dimensional model in the $ABAB$
gauge for a square lattice with
anisotropy $\alpha_D=v_F/v_\Delta=12$ along the $\Gamma$Y direction. The two
branches of each band correspond to $0 \leq k_y \leq \pi$ 
(solid curves) and $- 2 \pi \leq k_y \leq -\pi$ (dashed
curves). Only positive energy bands are plotted for clarity, negative
energy bands can be obtained through particle-hole symmetry. Energies
are in units of $E_1=\hbar v_F/d$.} 
\label{figa12}
\end{figure}

There are a number of consequences of the small overlap between
wavefunctions localized around different vortices, as the anisotropy
$\alpha_D$ grows larger. First of all, note that the one-dimensional 
Hamiltonian (\ref{H1D}) can be split in the following way
\begin{equation}
{\cal H}_{\mbox{\scr 1D}} = {\cal H}_0 + k_y \sigma_3
\end{equation}
where ${\cal H}_0$, the one-dimensional Hamiltonian computed at
$k_y=0$,  is independent of $k_y$ and $\sigma_3$ is the diagonal 
Pauli spin matrix. 
This implies that the expectation value of ${\cal H}_{\mbox{\scr
1D}}$ in the state $\psi(x)$, $\cal E=<{\cal H}_{\mbox{\scr
1D}}>_\psi$, obeys
\begin{equation}
\frac{\partial \cal E}{\partial k_y} = <\sigma_3>_\psi.
\end{equation}
If the wavefunction $\psi(x)$ is localized entirely around the left
(respectively right) vortex in the unit cell, its only non-zero
component will be the upper (resp. lower) one. In this case,
$<\sigma_3>_\psi = +1$ (resp. -1), and the slope of the energy bands is
fixed, regardless of the value of $\alpha_D$. This is the case for
lowest energy eigenstates close to the Y point, especially at large 
anisotropy, as the potentials that go into the one-dimensional Hamiltonian
(\ref{H1D}) correspond to a roughly triangular well for one component,
with the bottom of the well close to the energy eigenvalue, and an
almost constant potential for the other component, with a fairly large
energy difference between the constant potential and the low-energy
eigenvalue. In this approximation, the low-energy eigenstates of
${\cal H}_{\mbox{\scr 1D}}$ are Airy functions, for one component, and
zero for the other. In the repeated band scheme, we expect the band structure
close to the Y point to look like parallel lines, the positive energy
ones with slope +1, and the negative energy ones with slope -1 (the
negative slope curves with positive energy in Fig.~\ref{figa12}
correspond to values of $-2 \pi \le k_y \le - \pi$, in the extended
zone scheme). This is because around $k_y =\pi$, the potential
$U^A(x)+k_y$ is pushed up in energy, while $U^B(x)-k_y$ is pushed
down, and therefore, in the limit $\alpha_D \to \infty$, the positive
energy states are strongly localized around the left vortex in the
unit cell, while the negative energy states are localized around the
right hand one. 

Let us turn our attention to the first near crossing at approximately
zero energy along the $\Gamma$Y line, closest to the Y point. We claim
that the energy splitting between the first positive and negative
energy bands is exponentially small in the anisotropy $\alpha_D$. We
want to calculate the leading behavior of this splitting for large
$\alpha_D$, therefore we will work in the WKB approximation.
The leading exponential expression of the WKB wavefunctions
corresponding to energy $E$ takes the form
\begin{equation}
u(x) \sim e^{\pm \alpha_D \int^x \sqrt{-\left(E-k_y-U^A(x)\right)
\left(E+k_y-U^B(x)\right)}}
\end{equation}
and analogously for the lower component $v(x)$, as discussed in more
detail in Appendix A. 
We can calculate the lowest energy levels at the Y point using the
Bohr-Sommerfeld quantization rule. If $|E|< 2 \pi$, the turning points
for the left vortex (centered at $x=-1/4$) are $x^E_- = -\frac{1}{4} -
\frac{E}{2 \pi}$ and $x^E_+ = -1/4$, therefore the quantization
condition reads
\end{multicols}
\begin{equation}
\int_{x^E_-}^{x^E_+} dx\,  \sqrt{\left(E+2 \pi x +\frac{\pi}{2}\right)
\left(E+2 \pi x +\frac{3}{2} \pi \right)} = \frac{\pi}{\alpha_D}
\left(n-\frac{1}{4}\right),
\label{BScond}
\end{equation}
\begin{multicols}{2}
with $n \ge 1$. On the right hand side we have used the connection rules for a
soft boundary on one side and a hard boundary on the other side of the
classically allowed region, as we assumed that the energy of the
eigenstate is small compared to the height of the triangular well. 
The integral on the left hand side of equation (\ref{BScond}) can be
evaluated explicitly to find
\begin{equation}
\frac{1}{\pi}\left(E+\frac{\pi}{2}\right) \sqrt{E(E+\pi)} -
\frac{\pi}{8} \sinh^{-1} \sqrt{\frac{E}{\pi}} \sim \frac{1}{3 \sqrt{\pi}}
E^{3/2},
\end{equation}
where we have used $\sinh^{-1} x \sim x$ for $x \ll 1$. This condition
translates in our case to $E \ll \pi$, which, for large enough $\alpha_D$, is
satisfied even for large $n$, thereby justifying the usage of the 
Bohr-Sommerfeld quantization formula. In conclusion, we find the
quasiparticle spectrum at the Y point to be 
\begin{equation}
E_n \sim \pi \left(\frac{3
\left(n-\frac{1}{4}\right)}{\alpha_D}\right)^{2/3}
\end{equation}
with $n=1,2,\ldots$ From our previous discussion, we also know that as
we go away from the Y point the slope of the energy band is either +1
or -1, depending on their sign. This means that, if we neglect the
level repulsion for the moment, the lowest positive and negative
energy bands will cross at zero energy (because of the underlying
particle-hole symmetry) at $k_y^* \equiv \pi \left(1-\left(\frac{9}{4
\alpha_D}\right)^{2/3}\right)$. 

\begin{figure}
\noindent
\epsfxsize=8.5cm
\epsffile{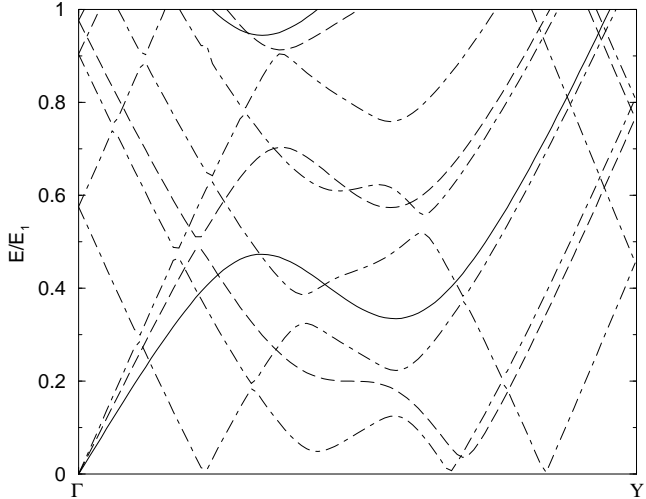}
\caption{Band structure of the one-dimensional model in the $ABAB$
gauge for a square lattice with
anisotropy $\alpha_D=v_F/v_\Delta=8, 15, 35$ along the $\Gamma$Y
direction (solid, dashed and dot-dashed curve respectively). Only
positive energy bands are plotted for clarity, negative
energy bands can be obtained through particle-hole symmetry.Energies
are in units of $E_1=\hbar v_F/d$.} 
\label{figGY}
\end{figure}

We can now find the magnitude of the splitting between the lowest
positive and negative energy bands at $k_y^*$. Our method follows closely
the LCAO approximation in molecular physics. We consider two
wavefunctions corresponding to the lowest-energy states localized
around each vortex for $k_y= k_y^*$, that is spinors with either a
non-vanishing upper or lower component. If these states were true eigenstates
of the one-dimensional Hamiltonian (\ref{H1D}), the corresponding
eigenvalue would be zero, by definition of $k_y^*$. Because the
``small'' component of the spinor wavefunction is not exactly zero, we
can evaluate variationally the exponential dependence of the lowest 
eigenvalues on the spatial separation of the wavefunctions calculating
the overlap integral between the WKB wavefunctions of the two zero-energy
states. Because of the periodicity of the potential, there are in
principle two regions which are classically forbidden and therefore
contribute to the tunneling amplitude between the wavefunction
localized around one vortex and the other. On the other hand, for
low-energy states, one of these regions (going from one vortex site to
the other, regardless of the value of the energy) is wider and the
product of the potentials is larger than in the other region,
therefore contributing only a subleading exponential behavior. For
this reason we will only be concerned with the region between $x_R^*
\equiv -1/4 - k_y^*/(2 \pi)$ and $x_L^* \equiv -3/4 + k_y^*/(2 \pi)$
and the overlap integral between the two wavefunctions is
\end{multicols}  
\begin{eqnarray}
\Delta_{\min} &\sim& \exp\left(-\alpha_D \int_{x_R^*}^{x_L^*} dx\,
\sqrt{-\left( -2 \pi x -\frac{3}{2} \pi + k_y^*\right) \left(-2 \pi x
- \frac{1}{2} \pi - k_y^*\right)}\right) \nonumber \\
&\sim& \exp \left[-\frac{\pi^2}{16}
\alpha_D \left(1-4\left(\frac{9}{4 \alpha_D}\right)^{2/3}\right)\right].
\end{eqnarray} 
\begin{multicols}{2}  
The first minimum away from the $\Gamma$ point along the $\Gamma$Y
line develops for $\alpha_D > 7$. For larger values of $\alpha_D$ more
minima in the band structure can be found (for example, see
Fig.~\ref{figGY}), increasing the density of
states close to zero energy, and those band gaps can be computed in an
analogous fashion to find more exponentially small energy splittings.

Another feature of the band structure that we can understand with this
simple model is the width of the lowest positive and negative energy
bands along the $\Gamma$X direction. Once again, to find a variational
estimate of this band width, we need
to compute the overlap integral between two low-energy wavefunctions. This
time, we can start from the two zero-energy states at the $\Gamma$
point and calculate the $\Gamma$X band assuming that it is a
narrow tight-binding band arising from the overlap of these two
states. Again, for large $\alpha_D$, we expect the width of this band
to be exponentially close to zero. The lowest positive and negative
energy bands along the $\Gamma$X direction in the Brillouin zone can
be calculated in the tight-binding approximation  
\begin{equation}
{\cal E}(k_x) = \pm \Delta \! E_{\Gamma X} \sin \frac{k_x}{2}
\label{eq:TB}
\end{equation}
where $\Delta \! E_{\Gamma X}$ is given by the overlap integral
of the two zero-energy wavefunctions. For large $\alpha_D$, we can
again use the WKB approximation to find the asymptotic leading
behavior of the bandwidth
\begin{equation} 
\Delta \! E_{\Gamma X} \sim e^{- \alpha_D \int_{-1/4}^{1/4} dx \,
\sqrt{-U^A(x) U^B(x)}} = e^{- \frac{\pi^2}{16} \alpha_D}.
\label{eq:DeltaEGX}
\end{equation} 

To summarize the asymptotic analysis of the one-dimensional model in
the vortex lattice geometry performed in this section, we have
explained the parallel
energy bands in the $\Gamma$Y direction close to the $\Gamma$ point and
their slope. We have also calculated the leading exponential dependence on
$\alpha_D$ of the $\Gamma$X bandwidth and of the minima of the $\Gamma$Y bands
close to zero energy. In the next section, we will describe the
numerical solution of the one-dimensional model and compare these
analytic results to the numerical data. These results are important to
understand the right crossover scale between the quantum mechanical
and semiclassical regimes of the density of states, as discussed in
the introduction and in our previous publication
\cite{marinelli00}. 

Very different results would be expected in a singular gauge which does not spatially
separate the low-energy wavefunctions, for example the $AABB$ gauge.
As we showed in the previous section, the
potentials $U^A(x)$ and $U^B(x)$ in the $AABB$ gauge are identical
(see eq.~(\ref{eq:AABB})), therefore the triangular potential wells
corresponding to a line of $A$ or $B$ vortices are not spatially
separated, as in the $ABAB$ case, and the wavefunctions localized
around the two vortices always have a finite overlap. The nature of
the asymptotic limit for large $\alpha_D$ is then very different from the
case discussed above. We do not expect to find an exponential
dependence of the features of the energy spectrum on the anisotropy,
in either one- or two-dimensions. Numerical calculations confirm this
claim, as will be shown in the following section.

\section{Numerical results}
\label{sect:numerics}
\subsection{Bravais vortex lattice: $ABAB$ gauge}
We have studied the one-dimensional Hamiltonian (\ref{H1D}) numerically
discretizing it on a real-space grid in one dimension. The reduced
dimensionality compared to the Bogoliubov--de Gennes operator
(\ref{Htransf}) allows for much smaller scale numerical calculations
and is the key to accessing the large-anisotropy regime. We choose a
real-space representation of the one-dimensional model
in order to be able to take advantage of sparse matrix diagonalization
algorithms (see Appendix B for further details on the numerical
methods), although we checked some of our results against a diagonalization
of the Hamiltonian (\ref{H1D}) in momentum space without finding any
appreciable difference. Knapp, Kallin and Berlinsky \cite{knapp00}
diagonalized the one-dimensional model in the $ABAB$ gauge 
in momentum space and did not find any appreciable difference from our
results in the range of overlap.

Let us start discussing the numerical diagonalization in the case of
the $ABAB$ singular gauge. From the analytical results discussed in
the previous section and our previous numerical investigation of the
two-dimensional Bogoliubov--de Gennes equation \cite{marinelli00}, we
expect to find points in the Brillouin zone where
the energy spectrum has an eigenvalue which is exponentially close to
zero. The simplicity of the numerical implementation of the
one-dimensional model allows us to resolve these eigenvalues in detail
and to compare them to our previous results in two dimensions.
  
Scanning the vortex lattice Brillouin zone we have computed the band
structure for several values of the anisotropy ratio $\alpha_D$. The
quasiparticle bands in a square vortex lattice are symmetric under the
exchange of $k_x \to -k_x$ or $k_y \to -k_y$, hence only positive
$k_x$ or $k_y$ have generally been considered. Also, particle-hole
symmetry holds at each 
point in the Brillouin zone, therefore only positive energies need to be
taken into account. In the one-dimensional model, the
potentials $U^{A,B}$ are periodic functions of $x$ and so the band
structure is 
periodic with respect to $k_x$, with period $2 \pi$. On the other hand, the
periodicity with respect to $k_y$ is lost, as can be seen, for example, in
Fig.~\ref{figa12} where the band structure of the one-dimensional model
$\vec{k} =(0,0)$ ($\Gamma$ point) and $\vec{k}=(0,\pi)$ (Y point) 
and the band structure between 
$\vec{k} =(0,-2 \pi)$ and $\vec{k}=(0,-\pi)$ (Y point) are plotted for
anisotropy $\alpha_D = 12$ respectively with a solid and dashed
curve. Note that the two branches, differing by a
reciprocal lattice vector of the two-dimensional vortex lattice
$\vec{Q} = (0,-2 \pi)$, cross at the $Y$ point and at non-symmetry
points. These crossings become avoided when the non-translational
invariance of the full two-dimensional Bogoliubov-de Gennes
Hamiltonian (\ref{Htransf}) is taken into account, but for large
enough anisotropy, the resulting gaps can be extremely small. Also,
these crossings and the changes they induce in the
density of states occur at finite energies and are not too crucial in the
understanding of the nature of the low-energy quasiparticle spectrum.    
\begin{figure}
\noindent
\epsfxsize=8.5cm
\epsffile{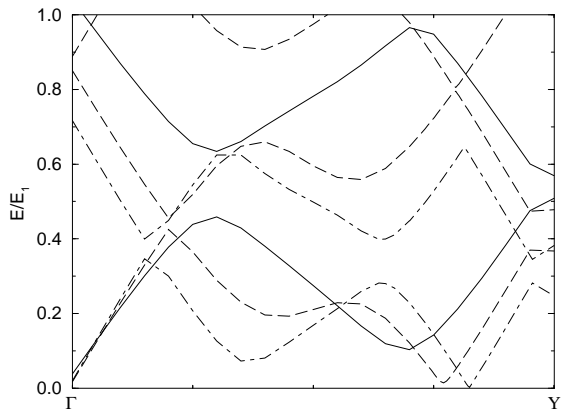}
\caption{Band structure of the two-dimensional linearized
Bogoliubov-de Gennes equation for a square lattice with
anisotropy ratio $\alpha_D=v_F/v_\Delta=8, 15, 35$ (solid, dashed and
dot-dashed respectively) along the $\Gamma$Y direction. Only positive
energy bands are plotted for clarity and negative
energy bands can be obtained through particle-hole symmetry. The gaps
at the $\Gamma$ point are fictitious, and are due to the chosen
lattice discretization scheme (Wilson fermions) of the Hamiltonian. 
In the $\alpha_D=15$  case, only the  two lowest
bands are plotted, while all the bands with energy $E<\hbar v_F/d$ are
plotted for $\alpha_D=8$ and 12. Energies are in units of $E_1=\hbar v_F/d$.} 
\label{figGY2D}
\end{figure} 

In Fig.~\ref{figGY} we have plotted the one-dimensional band structure for a square
vortex lattice with anisotropy ratio $\alpha_D =8, 15, 35$ along the
$\Gamma$Y axes. This plot should be compared to the two-dimensional
energy bands calculated in the same direction in the Brillouin zone
and plotted in Fig.~\ref{figGY2D}, previously published in
\cite{marinelli00}. The small gaps at the $\Gamma$ point in
Fig.~\ref{figGY2D} are fictitious and are are due to the
choice of real-space discretization (Wilson fermions \cite{wilson75})
of the linearized Bogoliubov--de Gennes equation in two dimensions. 
In one dimension, we use a different approach (staggered fermions \cite{kogut83}) 
which preserves the Dirac node at the $\Gamma$ point in the
discretized equations, described in more detail in
Appendix~\ref{app:numerics}. Similarly to what we observed for the two-dimensional
Hamiltonian (\ref{Htransf}), for large enough anisotropy, the energy
spectrum shows new low-energy states around a wavevector close to
the Y point. As we mentioned before, the degeneracies observed at the Y
point are split as soon as one introduces the non-translational
invariance of the Bogoliubov--de Gennes operator.
   
\begin{figure}
\noindent
\epsfxsize=8.5cm
\epsffile{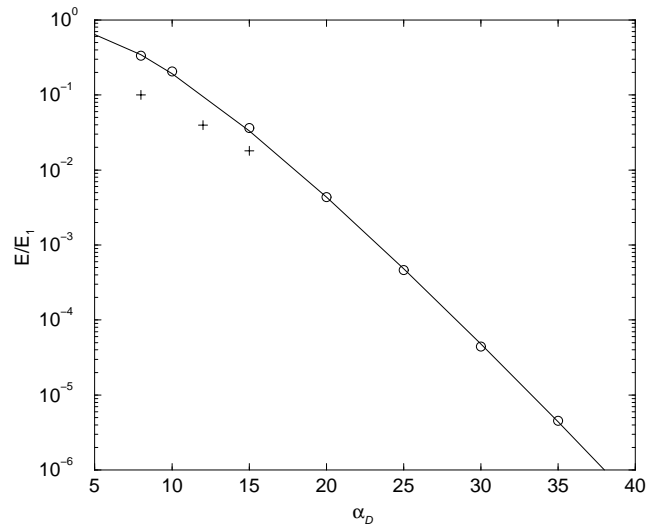}
\caption{Energy of the minimum of the positive lowest-energy band (away from
the $\Gamma$ point) along $\Gamma$Y as a function of anisotropy
$\alpha_D$. The circles ($\circ$) represent the results of the one-dimensional
model while the pluses (+) represent the two-dimensional one. The
solid curve is the asymptotic expression calculated in Sect.~III, with
the pre-exponential factor fitted to the one-dimensional
numerical data. Energies are in units of $E_1=\hbar v_F/d$.}  
\label{figmin} 
\end{figure}

The strong dependence of the minimum (away from the $\Gamma$ point) 
of the lowest-energy band along the $\Gamma$Y direction (closest to
the Y point) on the anisotropy $\alpha_D$ is plotted in
Fig.~\ref{figmin}, along with the asymptotic expression computed in
the previous section (the pre-exponential factor has been
fitted to the one-dimensional numerical data) and the
same data from a fully two-dimensional calculation for values of
$\alpha_D \le 15$. The accuracy of the asymptotic expression is
remarkably good even for relatively small values of $\alpha_D$ and the
two-dimensional data points are certainly consistent with the
one-dimensional approximation.

\begin{figure}
\noindent
\epsfxsize=8.5cm
\epsffile{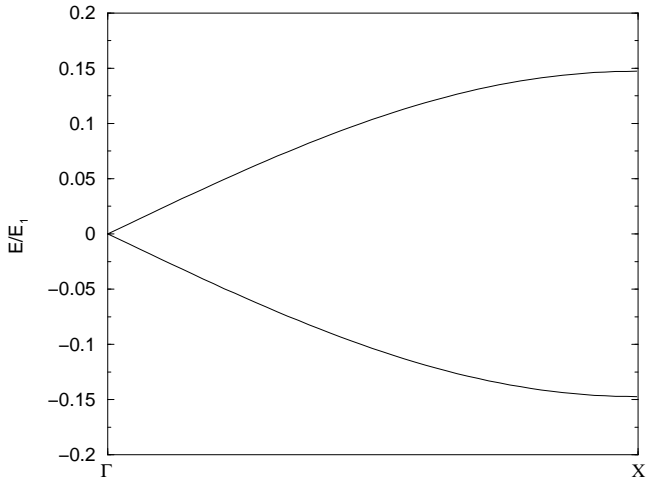}
\caption{Band structure for the one-dimensional model with
anisotropy $\alpha_D=v_F/v_/Delta=4$ along the $\Gamma$X direction. 
This curve is indistinguishable from the tight-binding expression 
$E = 0.15 \sin(k_x/2)$, plotted on the same scale. 
Energies are in units of $E_1=\hbar v_F/d$.}
\label{figGX}
\end{figure}

\begin{figure}
\noindent
\epsfxsize=8.5cm
\epsffile{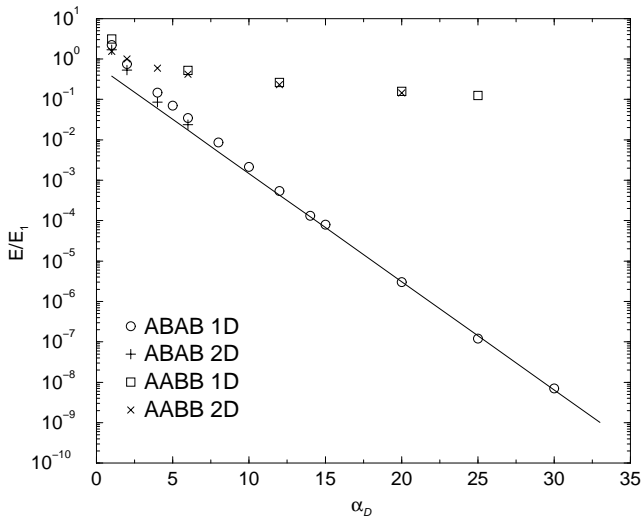}
\caption{Width of the lowest-energy band along $\Gamma$X as a
function of anisotropy $\alpha_D$. The solid curve represents the
asymptotic expression calculated in Eq.~(\protect{\ref{eq:DeltaEGX}}),
with the pre-exponential factor fitted to the one-dimensional
numerical data in the $ABAB$ gauge. 
The circles ($\circ$) are data points obtained by diagonalizing
numerically the one-dimensional model in the $ABAB$ gauge, 
while the pluses ($+$) show
the bandwidth in the $\Gamma$X direction of the two-dimensional
Hamiltonian also in the $ABAB$ gauge, for moderate anisotropy. The
last two data sets plotted show the same energy bandwidth for the one-
and two-dimensional model in the $AABB$ gauge, plotted with squares ($\Box$)
and crosses ($\times$), respectively. Energies are in units 
of $\hbar v_F /d$.}  
\label{figX}
\end{figure}

We have also computed the band structure in the $\Gamma$X direction 
and studied its dependence on the anisotropy ratio $\alpha_D$. 
Even for relatively modest values of the anisotropy, the lowest energy 
band is very well approximated by the tight-binding expression
\begin{equation}
{\cal E}(k_x) = \Delta\!E_{\Gamma X} \sin \frac{k_x}{2},
\end{equation}
as outlined in the previous section. In Fig.~\ref{figGX}, the band
structure of the one-dimensional model is plotted for anisotropy
$\alpha_D=4$, and it is indistinguishable from the tight-binding
expression above. We have computed numerically
$\Delta\!E_{\Gamma X}$ over a wide range of $\alpha_D$ and the results 
are shown in Fig.~\ref{figX}. The solid curve is the exponential
behavior computed in Sect.~\ref{sect:asymptotics}. The
asymptotic result agrees well with the numerical data for large
$\alpha_D$.

\subsection{Bravais vortex lattice: $AABB$ gauge}
Let us turn to the numerical diagonalization of the Bogoliubov--de
Gennes Hamiltonian for a square vortex lattice in the $AABB$ singular
gauge. Although it is still a Bravais lattice, the unit cell includes
four vortices, as shown in Fig.~\ref{fig:lattice} and it is a 1x2
rectangle, in dimensionless units. The coordinates of
the high-symmetry points in the first Brillouin zone are therefore
$\vec{k}=(0,0)$ ($\Gamma$ point), $\vec{k}=(\pi,0)$ (X point) and
$\vec{k}=(0,\pi/2)$ (L point). As we discussed in the previous
section, we do not
expect to find any exponentially small features in the energy band
structures for large values of the anisotropy ratio $\alpha_D$ in this
singular gauge. 

The different nature of the low-energy states in the two singular
gauges can be observed in Fig.~\ref{figX} where we plot the bandwidth
of the lowest $\Gamma$X energy band ($\Delta\!E_{\Gamma X}$, as
defined in equation (\ref{eq:TB}) in the previous section). Notice the
qualitatively different behavior in the two singular gauges we
considered. While the bandwidth in the $ABAB$ gauge follows very
closely the asymptotic exponential dependence on the anisotropy $\alpha_D$, 
the same energy scale in the $AABB$ gauge decays in a much slower
fashion and does not seem to reach an exponential decay in the
anisotropy range considered. In both cases, the one- and
two-dimensional calculations agree reasonably well with each other. 

We have studied both the one- and two-dimensional Hamiltonians in the
$AABB$ gauge. Even though the energy bands cannot be computed (even
just in the asymptotic limit of large anisotropy) by a simple approach 
as in the $ABAB$
case, the one-dimensional model gives some information about the more
involved two-dimensional Hamiltonian. In particular, the
one-dimensional calculation of energy bands in the region of the
Brillouin zone close to the $\Gamma$ point is in reasonably agreement
with the two-dimensional band structure. More generally, the
translational invariance in the direction of smearing of the
vortices implies that the different branches of the bands in the
$\Gamma$L direction --- differing by a reciprocal lattice vector
$\vec{Q}=(0,-\pi)$ --- can cross at the L point and at non-symmetry
points. These crossings, just like in the $ABAB$ gauge, become avoided
when the translational invariance is broken, as in the two-dimensional
case. Unlike in the $ABAB$ gauge, the magnitude of the resulting
energy gaps does not become exponentially small as we increase the
anisotropy ratio. This makes the one-dimensional model not as useful for
quantitative calculations of properties of the full two-dimensional
Bogoliubov--de Gennes Hamiltonian as in the $ABAB$ singular gauge.   

\begin{figure}
\noindent
\epsfxsize=8.5cm
\epsffile{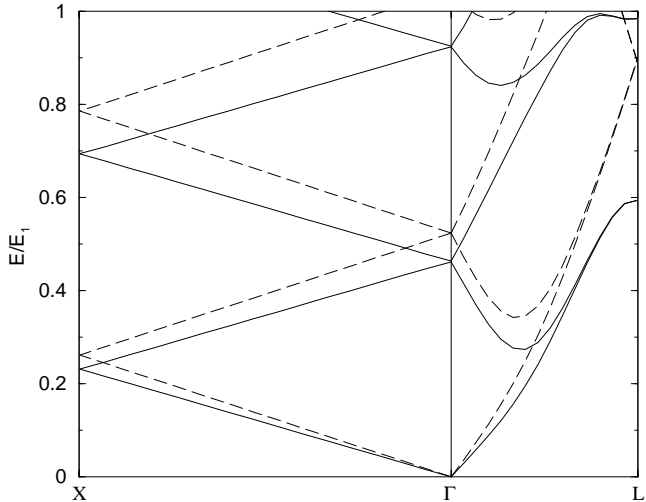}
\caption{Band structure for a square vortex lattice in the $AABB$
singular gauge with $\alpha_D=v_F/v_\Delta=12$. The solid line is the
spectrum of the two-dimensional Bogoliubov--de Gennes equation, while
the dashed line represents the one-dimensional model. Only positive
energy bands are plotted for clarity, negative energy ones can be
obtained by particle hole symmetry. Energies are in units of $E_1 =
\hbar v_F/d$.}  
\label{fig:AABB12}
\end{figure}

\begin{figure}
\noindent
\epsfxsize=8.5cm
\epsffile{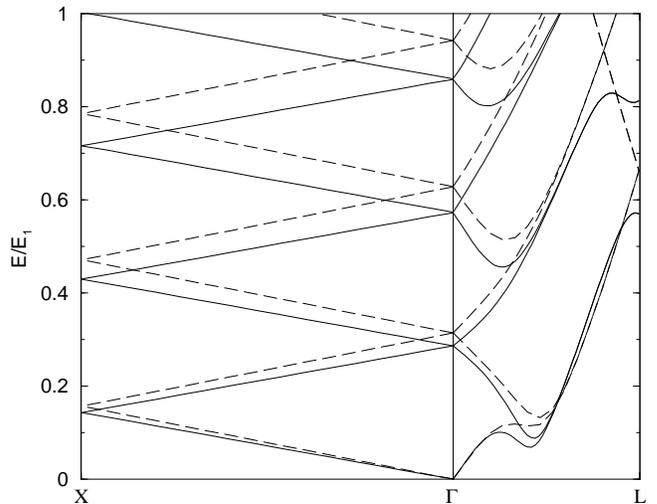}
\caption{Band structure for a square vortex lattice in the $AABB$
singular gauge with $\alpha_D=v_F/v_\Delta=20$. The solid line is the
spectrum of the two-dimensional Bogoliubov--de Gennes equation, while
the dashed line represents the one-dimensional model. Only positive
energy bands are plotted for clarity, negative energy ones can be
obtained by particle hole symmetry. Energies are in units of $E_1 =
\hbar v_F/d$.}  
\label{fig:AABB20}
\end{figure}

The one- and two-dimensional band structures are plotted in
Figs.~\ref{fig:AABB12} and \ref{fig:AABB20}. The energy bands are
computed in the $\Gamma$X and $\Gamma$L directions for anisotropies
$\alpha_D=12$ and 20.  Although the one-dimensional model
fails to reproduce all the details of the two-dimensional band
structure away from the $\Gamma$ point even for relatively large
values of the anisotropy, the one-dimensional bands allow to sketch
the qualitative behavior of the two-dimensional ones, once the band
crossings become avoided. Finally, notice that, while in the $ABAB$ 
gauge there was a minimum developing in the 
lowest energy band already for $\alpha_D =8$, in the $AABB$ gauge the
anisotropy ratio has to be much larger ($\alpha_D=20$) before we start
finding an inflection in the lowest energy band for both the one- and
two-dimensional Bogoliubov--de Gennes operators. 
 
In conclusion, even though the band structures in the $ABAB$ and
$AABB$ gauges are qualitatively similar close to the center of the
Brillouin zone, they differ significantly near the edges of the Brillouin
zone. While in the $ABAB$ case all relevant energy scales depend
exponentially on the anisotropy ratio $\alpha_D$, the same is not true
for the $AABB$ gauge. This behavior is well understood in terms of the
simplified one-dimensional model in both gauges, for large values of
the anisotropy ratio $\alpha_D$.

\subsection{Non-Bravais vortex lattice}
\begin{figure}
\noindent
\epsfxsize=8.5cm
\epsffile{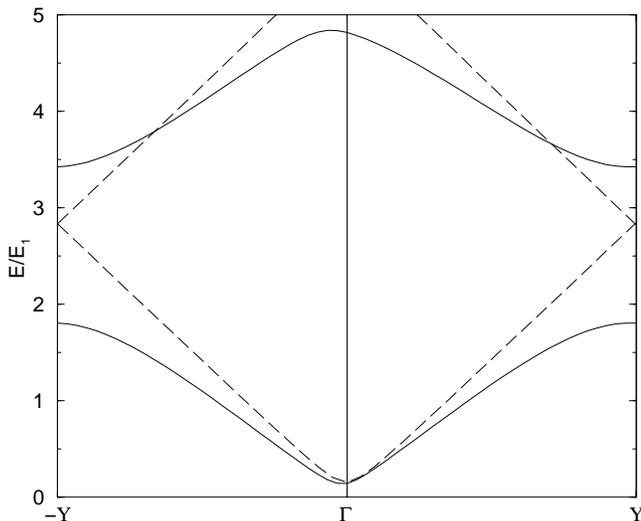}
\caption{Band structure for a non-Bravais square vortex lattice in the $ABAB$
singular gauge from -Y to Y with $\alpha_D=v_F/v_\Delta=1$ and vortex
coordinates in the unit cell $(\pm 1/5,\pm 1/5)$. The solid line is the
spectrum of the two-dimensional Bogoliubov--de Gennes equation, while
the dashed line represents the one-dimensional model. Only positive
energy bands are plotted for clarity, negative energy ones can be
obtained by particle hole symmetry. Energies are in units of $E_1 =
\hbar v_F/d$.}  
\label{fig:nonbravGY1}
\end{figure}

\begin{figure}
\noindent
\epsfxsize=8.5cm
\epsffile{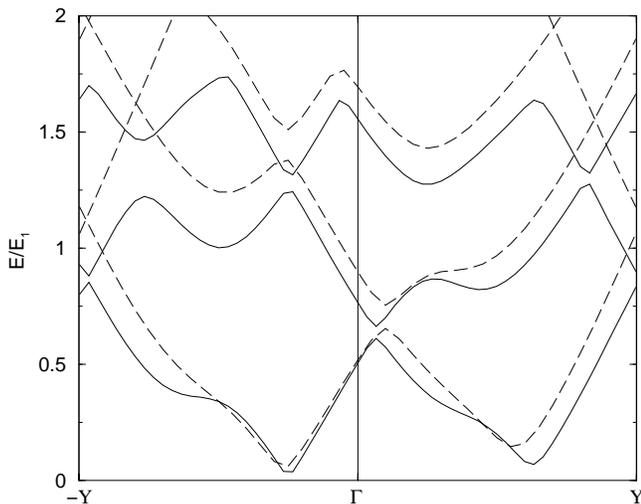}
\caption{Band structure for a non-Bravais square vortex lattice in the $ABAB$
singular gauge from -Y to Y with $\alpha_D=v_F/v_\Delta=8$ and vortex
coordinates in the unit cell $(\pm 1/5,\pm 1/5)$. The solid line is the
spectrum of the two-dimensional Bogoliubov--de Gennes equation, while
the dashed line represents the one-dimensional model. Only positive
energy bands are plotted for clarity, negative energy ones can be
obtained by particle hole symmetry. Energies are in units of $E_1 =
\hbar v_F/d$.}  
\label{fig:nonbravGY8}
\end{figure}

The Bravais lattice nature of the vortex lattices we considered so far
ensures the gaplessness of the spectrum at the center of the
Bruillouin zone \cite{marinelli00}. In the $ABAB$ singular gauge, we
can consider a unit cell with two vortices again, but relax the
constraint that they sit equidistantly from each other, along the
diagonal --- \emph{i.e.} the magnitude of $\vec{R_0}$, as defined in
Fig.~\ref{figAB}, does not need to be equal to $\sqrt{2}d/4$. Let us
assume that the two sublattices $A$ and $B$ are still square lattices
with spacing $d$, but let us consider, for concreteness, the case
where the distance between nearest $A$ and $B$ vortices is
$(2\sqrt{2}/5)d$, which corresponds to vortex coordinates in the unit
cell of $(\pm 1/5,\pm 1/5)$ in dimensionless units. We can use the
one-dimensional potentials computed in equations (\ref{eq:xvA}) and
(\ref{eq:xvB}) with $x_v=1/5$ to determine the one-dimensional band 
structure. The energy bands along the -Y to Y direction for anisotropy
ratios $\alpha_D=1$ and 8 are plotted in Figs.~\ref{fig:nonbravGY1}
and \ref{fig:nonbravGY8}, respectively. First of all, we notice that
the $k_x \to -k_x$ and $k_y \to -k_y$ symmetries of the energy
spectrum are broken. Fig.~\ref{fig:nonbravGY8} exemplifies how the
one-dimensional very effectively captures the qualitative features of
the two-dimensional band structure also for a non-Bravais lattice in
the large anisotropy limit for the same reasons discussed above in the
Bravais lattice case.        

We want to emphasize here that, unlike for a Bravais lattice of
vortices, the energy spectrum here is gapped. These gaps where
discussed in a perturbation theory framework in
\cite{marinelli00}. What is interesting to note is that the evaluation
of these gaps in the one-dimensional model is very consistent with the
two-dimensional calculation, even though the handling of the
discretization of the linearized Bogoliubov--de Gennes equation is
different in the one- and two-dimensional case, as is explained in
more detail in Appendix B. 

\section{Conclusions}
\label{sect:conclusion}
We have studied the large Fermi-velocity anisotropy $\alpha_D =
v_F/v_\Delta$ limit of the quasiparticle spectrum of a $d-$wave
superconductor in the mixed state. The vortex geometry we considered
is that of a square vortex lattice rotated by $45^\circ$ from the
anisotropy axes. The original linearized Bogoliubov--de Gennes
equation was mapped onto a Dirac Hamiltonian in an effective periodic
vector and scalar potential using the 
Franz--Te\v{s}anovi\'c gauge transformation. We have
studied the dependence of the energy spectrum on the choice of such
singular gauge transformation. Physically, one might expect the gauge
choice not to matter. However, although the Bogoliubov--de Gennes operator
considered in this paper is indeed gauge invariant away from the
vortices, the singularity at the vortex site introduces boundary
conditions that break the invariance of the energy spectrum on the
choice of singular gauge.

In the large anisotropy limit, we have solved a one-dimensional
variant of the Franz--Te\v{s}anovi\'{c} model obtained by smearing the
vortices in the ``hard'' direction. We computed the one-dimensional 
potentials associated with two different choices of singular gauges 
in the case of a Bravais square
vortex lattice. In one case (the $ABAB$ gauge), we found that the
low-energy localized states around each vortex type ($A$ or $B$) in
the one-dimensional potential wells are physically separated. The
model can therefore be solved in the WKB approximation and several
features of the energy band structure are shown to depend
exponentially on the anisotropy ratio $\alpha_D$. We compared the
computed energy spectrum to the results of a numerical diagonalization of the
Hamiltonian (\ref{H1D}). In both cases we found that both the minima
of the lowest energy-band in the $\Gamma$Y direction and the bandwidth
of the lowest energy-band in the $\Gamma$X direction decay exponentially to
zero as a function of the anisotropy $\alpha_D$. The asymptotic
calculation agrees very well with the numerical results. 

Different results are obtained in the $AABB$ gauge. In this case, the
low-energy states localized around the vortices of different types are
not spatially separated. This results in a finite overlap integral for
any value of the anisotropy and, therefore, no exponentially small
features in the band structure. The one- and two-dimensional numerical
spectra in this gauge are not as similar as in the $ABAB$ gauge, but
one can still extract information like degeneracies or other global
properties of the two-dimensional band structure from the
one-dimensional calculations. Of course, the detailed spectra will not
be equal even in the asymptotic large $\alpha_D$ limit.

Finally, we have distorted the square vortex lattice in the $ABAB$
gauge into a non-Bravais vortex lattice. Again, we find good agreement
between the one- and two-dimensional calculations of the energy
spectra. In particular, we find a gap in the lowest energy band close
to the $\Gamma$ point (exactly at the center of the Brillouin zone
only in the $\alpha_D=1$ case, otherwise the $k_x \to -k_x$ and $k_y \to
-k_y$ symmetry of the energy bands is broken) in both numerical
calculations. 

\section{Acknowledgement}
The authors are grateful for helpful discussions with S.\ Simon, M.\ Hermele,
Z.\ Te\v{s}anovi\'{c}, M.\ Franz, A.\ Vishwanath, and E.\ Demler. This
work was supported in part by NSF grant DMR-99-81283.

\appendix

\section{WKB approximation for the one-dimensional model}
\label{app:WKB}
We want to find the eigenfunctions for a given energy $E$ of the
one-dimensional model in the WKB approximation and derive the
appropriate Bohr-Sommerfeld quantization condition. Starting from the
one-dimensional Hamiltonian (\ref{H1D}), let us define the potentials
$\Phi^A(x) \equiv k_y + U^A(x)$ and $\Phi^B(x) \equiv -k_y +U^B(x)$
to simplify notations. If we fix the energy $E$, we need to find a
solution to the system of differential equations
\begin{equation}
\left\{
\begin{array}{c}
\Phi^A(x) u(x) + \left(\frac{1}{i \alpha_D} \partial_x +
\frac{1}{\alpha_D} k_x\right) v(x) = E u(x) \\
\left(\frac{1}{i \alpha_D} \partial_x +
\frac{1}{\alpha_D} k_x\right) u(x) + \Phi^B(x) v(x) = E v(x).
\end{array}
\right.
\label{eqWKB}
\end{equation}
The derivation of the WKB form of the wavefunctions proceeds along
standard lines (see for example \cite{bender78}). Substituting
the expansion
\begin{equation}
\begin{array}{c}
u(x) = e^{i \alpha_D \sum_{n=0}^{\infty} S_n(x) \alpha_D^{-n}} \\
v(x) = e^{i \alpha_D \sum_{n=0}^{\infty} T_n(x) \alpha_D^{-n}} 
\end{array}
\label{eqexpWKB}
\end{equation}
in the previous system of equations, to find the asymptotic behavior
of the wavefunctions for large $\alpha_D$. Expanding and equating
order by order in $\alpha_D$ in equation (\ref{eqWKB}), we can
calculate the first two terms in the series (\ref{eqexpWKB}), namely
$S_{0,1}(x)$ and $T_{0,1}(x)$. These are the leading terms in the
expansion for large $\alpha_D$ and the truncation of the series
(\ref{eqexpWKB}) to this order is the WKB approximation.
In particular, $S_0(x)$ and $T_0(x)$ turn out to be 
\begin{equation}
S_0(x) = T_0(x) = \pm \int^x dx' \, \sqrt{(E-\Phi^A(x'))(E-\Phi^B(x'))}
\end{equation}
which gives the leading exponential behavior and is the key expression for
the Bohr-Sommerfeld quantization formula. Every other term in the series
expansion leads to algebraic corrections, which we have not taken into
account in this work. For completeness, we state the expression of the
WKB wavefunctions, including the algebraic prefactor:
\end{multicols}
\begin{equation}
\begin{array}{c}
u(x) \sim \left(\frac{E-\Phi^B}{E-\Phi^A}\right)^{1/4} e^{\pm i k_x x}
e^{\pm i
\alpha_D \int^x dx' \, \sqrt{(E-\Phi^A(x'))(E-\Phi^B(x'))}} \\
v(x) \sim \left(\frac{E-\Phi^A}{E-\Phi^B}\right)^{1/4} e^{\pm i k_x x}
e^{\pm i
\alpha_D \int^x dx' \, \sqrt{(E-\Phi^A(x'))(E-\Phi^B(x'))}}.  
\end{array}
\end{equation}
\begin{multicols}{2}
Obviously, these oscillating expressions are correct in the
classically allowed regions, while the exponential factors become real
in the classically forbidden regions.

\section{Numerical Implementation of the One-dimensional Model}
\label{app:numerics}
To diagonalize the Hamiltonian (\ref{H1D}) we have used a real-space
representation to take advantage of fast sparse-matrix diagonalization
algorithms and to control the smoothing of vortex cores over a finite
region of space, which cannot be avoided in a momentum-space
representation. The main drawback of discretizing a Dirac-type
Hamiltonian in real space is the appearance of the ``Fermion doubling
problem'' (see \cite{kogut83} for a thorough discussion of possible
ways of discretizing a Dirac Hamiltonian). 
The real-space representation of a Dirac Hamiltonian leads to the 
introduction of $2^D -1$ spurious large-wavevector, low-energy modes 
(additional Fermions, hence the name), where $D$ is the dimensionality of
the system. In our case, the system being one dimensional, there is
only one spurious mode, which we can eliminate using the staggered-fermion
approach \cite{kogut83}. If we discretize the vortex lattice unit cell
with a mesh of size $h = 1/(N-1)$
and index the resulting mesh with an integer $n$ ranging from $-N/2$
to $N/2$ so that $x = n/N$, we can place a
single-component Fermi field $\xi(n)$ on each 
site. To identify a single two-component Dirac field, we can decompose
the mesh into an even and an odd sublattice and define
\begin{equation}
\begin{array}{cc}
V(n) = \xi(n), & n \mbox{ even} \\
W(n) = \xi(n), & n \mbox{ odd}.
\end{array}
\end{equation}
The staggered fermion method avoids the Fermion doubling problem by
doubling the size of the unit cell in real space, thus
halving the size of the Brillouin zone (and overall number of degrees
of freedom). Note that the relevant unit cell here is unrelated to the
vortex lattice unit cell, it is simply the unit element of the mesh
with which we discretize the continuum equations. Defining the function 
\begin{equation}
G(n) = \left\{ 
\begin{array}{cc}
U^A(n) + k_y, & n \mbox{ even} \\
U^B(n) - k_y, & n \mbox{ odd}
\end{array} \right.
\end{equation}
the discretized eigenvalue problem takes the form
\end{multicols}
\begin{equation}
\frac{1}{i \alpha_D} \frac{\xi(n+1)-\xi(n-1)}{2 h} + G(n) \xi(n)
+ \frac{1}{\alpha_D} k_x \frac{\xi(n+1) + \xi(n-1)}{2} = E \xi(n)
\end{equation}
\begin{multicols}{2}
and the correct continuum limit is achieved by letting $h \to 0$,
without any further caveats. An important advantage of 
staggered fermions over Wilson fermions \cite{wilson75}
(which we used previously \cite{marinelli00}) is that Wilson approach
breaks the chiral symmetry of the translationally invariant 
Hamiltonian,
introducing a $k-$dependent mass term to lift the spurious modes at
the edges of the mesh Brillouin zone to high energy. By explicitly
breaking the chiral symmetry, fictitious gaps can and do appear in the
energy spectrum at the center of the vortex lattice Brillouin zone
thus making it difficult to perform a precise study of almost
dispersionless bands, which is one of the goals of the present work. 
In principle, these gaps could be eliminated to arbitrary precision by 
carefully tuning suitable counterterms but the procedure (having to be
performed numerically) is slow and not very accurate. The
staggered fermion approach bypasses all these difficulties by
preserving explicitly the chiral symmetry of the one-dimensional
Hamiltonian in the lattice representation, if there was one initially
in the continuum representation.


\end{multicols}
\end{document}